\documentclass[twocolumn]{jpsj3} 
\usepackage{bm}
\usepackage{color}

\title{Magnetic Response in Quantized Spin Hall Phase of Correlated Electrons}

\author{Jun \textsc{Goryo}
\thanks{E-mail address: jungoryo@iis.u-tokyo.ac.jp} and 
Nobuki \textsc{Maeda}$^{1}$
\thanks{ E-mail address: maeda@particle.sci.hokudai.ac.jp}}

\inst{Institute of Industrial Science, the University of Tokyo, 4-6-1 Komaba, Meguro, Tokyo, Japan \\
$^{1}$ Department of Physics, Hokkaido University, Sapporo, Japan}

\kword{magnetic response, quantized spin Hall effect, Kane-Mele model, intrinsic spin-orbit interaction, electron correlation, layered-honeycomb structure, superlattice, 
topological BF term, superconductivity}

\abst{
We investigate the magnetic response in the quantized spin Hall (SH) phase of a layered-honeycomb lattice system 
with intrinsic spin-orbit coupling $\lambda_{\rm SO}$ and on-site Hubbard $U$. 
The response is characterized by the parameter $g= 4 U a^2 d / 3$, 
where $a$ and $d$ are the lattice constant and interlayer distance, respectively. 
When $g< (\sigma_{xy}^{s2}  \mu)^{-1}$, where $\sigma_{xy}^{s}$ is 
the quantized spin Hall conductivity and $\mu$ is the magnetic permeability, 
the magnetic field inside the sample oscillates spatially. The oscillation vanishes in the non-interacting limit $U\rightarrow 0$. 
When $g > (\sigma_{xy}^{s2} \mu)^{-1}$,  the system shows perfect diamagnetism, i.e., the Meissner effect occurs. 
We find that a superlattice structure with large $a$ is favorable for observing these phenomena. 
We also point out that, as a result of Zeeman coupling, the topologically protected helical edge states 
show weak diamagnetism that is independent of $g$. }

\begin{document}
\maketitle

\section{Introduction}

The aim of this paper is to discuss the magnetic response of the quantized spin Hall (SH) phase of correlated electrons on 
the layered-honeycomb lattice.

It has been shown that the quantum SH effect occurs in a non-interacting 
electron system with intrinsic spin-orbit coupling $\lambda_{\rm SO}$, like the Kane-Mele (KM) model.\cite{KM1,KM2,note2}   
The investigation of the quantum SH system with on-site Hubbard $U$ is now 
one of the current topics, \cite{Shitade,Pesin-Balents,Rachel-LeHur} 
and the phase diagram has been obtained.\cite{Pesin-Balents,Rachel-LeHur} In ref. 7, magnetic response 
in the quantized SH phase of 
the layered KM model with $U$ (spin-conserving limit of the topological band insulating phase in ref. 6) was  discussed. 
Such a model can be applied to correlated electrons in the system with honeycomb layers, such as some Ir-based oxides.\cite{Shitade} 
The correlation is characterized by the parameter $g \propto U a^2 d$ in the low-energy long-wavelength regime 
($a$; lattice constant, $d$; interlayer distance), and the London equation for the Meissner effect  
has been obtained in the large $g$ limit.\cite{GM} The discussion on general $g$ has not yet been given. 
The role of the topologically protected helical edge states, which is the hallmark of the quantum SH effect,\cite{KM1,KM2,HgTe-th,HgTe-exp,Hasan-Kane} 
has also not been taken into account. 

In this study, we clarify the magnetic response for general $g$ in the quantized SH phase. 
First, we take into account contributions from bulk states and later consider those from the helical edge states.  
When $g < (\sigma_{xy}^{s2}  \mu)^{-1}$, where $\sigma_{xy}^s$ and $\mu$ are the quantized spin Hall conductivity (SHC) and magnetic permeability, respectively, 
the magnetic field inside the sample oscillates spatially around a constant value. 
The oscillation vanishes in the non-interacting limit $U\rightarrow 0$. 
When $g > (\sigma_{xy}^{s2} \mu)^{-1}$, the general solution for the magnetic field becomes a superposition of 
the homogeneous and damping parts.  We find that the damping part is energetically favored, thus, the Meissner effect occurs.
Then, we consider the contribution from the helical edge state. As a result of Zeeman coupling, it is shown that the state exhibits weak 
diamagnetism that is independent of $g$. 

The argument we will present is focused on the quantized SH phase. 
Here, we mention that we have an upper limit for $g (\propto U a^2 d)$. It has been pointed out 
that the system shows a phase transition from the quantized SH phase to the topological Mott insulating phase 
when $U$ becomes larger.\cite{Pesin-Balents,Rachel-LeHur,note0} The lattice constant $a$ also should not be too large, since 
we consider the long-wavelength effective theory.  

This paper is organized as follows. 
In $\S$ 2, we introduce the layered KM model with on-site Hubbard $U$. 
In $\S$ 3, we integrate out Fermion and obtain a one-loop effective Lagrangian in the quantized SH phase. 
In $\S$ 4, we discuss the magnetic response. 
In $\S$ 5, we estimate the contribution from the helical edge state. 
In $\S$ 6, we comment on 
the relation to the superconductivity and our system. 
We use $\hbar=c=1$ unit and the Minkovskian metric $g^{\mu\nu}=diag(1,-1,-1)$, where $\mu,\nu=0,x,y$. Summations run 
over repeated Greek indices. 

\section{Layered KM Model with an Electron Correlation}
We consider electrons on the layered honeycomb lattice.  We assume that 
interlayer coupling is negligibly small and each layer is described by the KM model.\cite{Rachel-LeHur,GM}
One of the essential ingredients of the KM model\cite{KM1,KM2} is the intrinsic SO coupling $\lambda_{\rm SO}$. We can say, not strictly but intuitively, that 
$\lambda_{\rm SO}$ gives an effective magnetic field depending on spin. It also gives an electronic excitation gap,\cite{KM2}
\begin{eqnarray}
\Delta=3\sqrt{3} \lambda_{\rm SO}. 
\label{SO-gap}
\end{eqnarray}
Thus, as an analog of the quantum Hall effect, we see quantization of the SHC\cite{KM1,KM2}, 
\begin{eqnarray}
\sigma_{xy}^s=\frac{e}{2\pi d}\frac{\Delta}{|\Delta|}, 
\label{SHC}
\end{eqnarray}
where $d$ is the interlayer distance. The model can also have the Rashba extrinsic SO coupling $\lambda_R$, which 
breaks the inversion symmetry and is induced by an electric field perpendicular to the honeycomb lattice  plane. The term also breaks the conservation of electron spin $S_z$. Hereafter, we consider the case of $\lambda_R=0$.  

We add the on-site Coulomb repulsion $U>0$. The Hamiltonian per layer is 
\begin{eqnarray}
H&=&t\sum_{<ij>}c^\dagger_i c_j + i \lambda_{\rm SO} \sum_{<<ij>>}\nu_{ij} c^{\dagger}_i s_z c_j 
\nonumber\\
&&+ U\sum_i n_{i \uparrow}n_{i \downarrow},  
\label{latticeH}
\end{eqnarray}
where $c_i$ ($c^\dagger_i$) is the annihilation (creation) operator of an electron with spin at the $i$-th site and $t$ is the nearest neighbor hopping. The second term is the intrinsic SO term consisting of the next nearest neighbor hopping, and 
$\nu_{ij}=\frac{2}{\sqrt{3}} (\hat{\bm d}_1\times \hat{\bm d}_2)_z=\pm1$, where  $\hat{\bm d}_1$ and  $\hat{\bm d}_2$ are unit vectors along the two bonds 
where the electron moving from site $j$ to $i$ passes.

Let us discuss how to deal with the electron correlation $U$. 
On-site Coulomb repulsion can be written by the on-site spin-spin interaction 
\begin{eqnarray}
U n_{i\uparrow}n_{i\downarrow}=\frac{U}{2}(n_{i\uparrow}+n_{i\downarrow}) - \frac{U}{6}(c^\dagger_i \vec{s} c_i)^2. 
\end{eqnarray}
The first term in the {\it r.h.s} merely gives the renormalization for the chemical potential and can be neglected. 
We introduce the auxiliary field $\vec{\varphi}_i$, which is a three-component vector in the spin space, 
and use the Stratonovich-Hubbard transformation,\cite{Mahan} 
$H \rightarrow H_{SH}=H+\Delta H,$ where 
\begin{eqnarray}
\Delta H&=&\frac{U}{6}\sum_i (c^\dagger_i \vec{s} c_i - \frac{3}{2U} \vec{\varphi}_i)^2, 
\label{SHgauss}
\\
H_{SH}&=&t\sum_{<ij>}c^\dagger_i c_j + i \lambda_{\rm SO} \sum_{<<ij>>}\nu_{ij} c^{\dagger}_i s_z c_j 
\nonumber\\
&&-\sum_i \vec{\varphi}_i \cdot c_i^\dagger \frac{\vec{s}}{2} c_i + \frac{3}{8U}\sum_i |\vec{\varphi}_i|^2. 
\label{latticeHSH}
\end{eqnarray}
The spin-spin interaction is eliminated in appearance, but instead, we have coupling between $\vec{\varphi}_i$ and  the electron spin, and a quadratic term of $\vec{\varphi}_i$.

We consider the continuum limit and take into account the low-energy electronic 
excitations around $K$ and $K^\prime$ points in the Brillouin Zone,\cite{KM1,KM2} i.e., we omit the inter valley scattering. 
We introduce the electromagnetic $U(1)$ gauge field 
$A_\mu$ and $SU(2)$ spin gauge field $\vec{a}_\mu$ via the 
covariant derivative 
\begin{eqnarray}
i D_\mu= i \partial_\mu - e A_\mu + \vec{a}_\mu \cdot \frac{\vec{s}}{2},  
\end{eqnarray}
where $\vec{a}_0=\vec{\varphi}$ (the auxiliary field in the continuum limit) and $\vec{\bm a}$ is an external field introduced artificially to estimate the spin current. 
We define a parameter 
\begin{eqnarray}
g=\frac{4 U a^2d}{3}, 
\label{g}
\end{eqnarray}
where $a$ is the lattice constant, and the microscopic Lagrangian density is\cite{note}  
\begin{eqnarray}
{\cal{L}}&=&\Psi^\dagger \left\{i D_0 - iv (D_x \tau_z \sigma_x + D_y \sigma_y) 
+ \Delta \tau_z \sigma_z s_z \right\}\Psi
\nonumber\\
&&
+ \frac{\epsilon_0 E^2}{2}- \frac{B^2}{2\mu_0} -\frac{1}{2g}|\vec{a_0}|^{2},   
\label{microL}
\end{eqnarray}
where $\Psi=\Psi_{\tau\sigma s}$ is the eight-component fermion field labeled by the eigenvalues of the diagonal components of valley spin $\vec{\tau}$, sublattice spin $\vec{\sigma}$, and real spin $\vec{s}/2$. The parameter $v$ is the Fermi velocity when the system is in the metallic state, and $\epsilon_0$ and $\mu_0$ denote the vacuum values of the dielectric constant and magnetic permeability, respectively.  
Note that, except for the last term, the Lagrangian (\ref{microL}) possesses the $U(1)_{\rm em} \times U(1)_z$ local gauge symmetry.  The $SU(2)$ gauge symmetry is broken down to $U(1)_z$, since the SO term contains $s_z$.

\section{One-Loop Effective Lagrangian in the Quantized SH Phase}

The phase diagram for correlated electrons on pyroclore and honeycomb lattices with $\lambda_{\rm SO}$ 
has been discussed using the slave-rotor model.\cite{Pesin-Balents,Rachel-LeHur} The system with strong $\lambda_{\rm SO}$ and small $U$ is in the quantized SH phase when the spin conservation is preserved. As $U$ increases, the band gap closes and the Mott gap opens instead, i.e., band-Mott transition occurs. We focus on the quantized SH phase.  

The derivation of the effective action in this phase is equivalent to that presented in ref. 14, although the physical meaning of the spin gauge field is different. We integrate out $\Psi$ from eq. (\ref{microL}) and obtain the one-loop effective Lagrangian for the gauge fields. 
We assume that the amplitude of the gauge fields is small and use the Gaussian approximation. We consider fields with a length scale (temporal scale) of modulation that is sufficiently gradual compared with $a$ ($\Delta$), and use the long-wavelength (low-frequency) approximation.

The result is\cite{GMI}
\begin{eqnarray}
{\cal{L}}_{\rm eff}&=&-\frac{1}{2g}a_0^{z2}+{\cal{L}}_{\rm ind}, 
\label{Leff}
\\
{\cal{L}}_{\rm ind}&=&\sigma_{xy}^s \epsilon^{\mu\rho\nu} a_\mu^z \partial_\rho A_\nu 
+ \frac{\epsilon E^2}{2} - \frac{B^2}{2\mu}+\frac{\epsilon_s e_s^2}{2}- \frac{b_s^2}{2 \mu_s}
\nonumber\\
&&+ ({\rm terms~independent~of}~a_\mu^z~{\rm and}~A_\mu),  
\label{Lind}
\end{eqnarray}
where ${\cal{L}}_{\rm ind}$ stands for the induced part of the effective Lagrangian, $\epsilon^{012}=\epsilon^{120}=\epsilon^{201}=-\epsilon^{021}=-\epsilon^{210}=-\epsilon^{102}=1$, and ${\bm e}_s=-\dot{\bm a}^z -{\bm \nabla} a_0^z$ and $b_s=\sum_{ij} \epsilon^{ij} \partial_i a_j^z$ with $\epsilon^{12}=-\epsilon^{21}=1$ are the spin electric field and spin magnetic field, respectively. The first term in eq. (\ref{Lind}) is the topological 
BF term,\cite{GM,GMI,BF1,BF2,BF3,BF4}
which plays an important role in our discussion. The coefficient is the quantized SHC given in eq. (\ref{SHC}). 
Note that only $a_\mu^z$ couples to the electromagnetic gauge fields. This comes 
from the fact that the $SU$(2) symmetry is broken down to $U(1)_z$ symmetry by the SO coupling.  We comment on 
the more detailed properties of this term in the Appendix. The Maxwell term is renormalized 
as\cite{GMI,Semenoff-Sodano-Wu}
\begin{eqnarray}
\epsilon&=&\epsilon_0+\delta \epsilon, 
\label{epsilon}\\
\frac{1}{\mu}&=&\frac{1}{\mu_0}+\frac{1}{\delta \mu}, 
\label{mu}\\
\delta \epsilon&\equiv& \frac{e^2}{6 \pi |\Delta| d}, 
\label{delta_epsilon}\\
\frac{1}{\delta \mu}&\equiv& \frac{e^2 v^2}{6 \pi |\Delta| d}, 
\label{delta_mu}\\
\epsilon_s&\equiv&\frac{\delta \epsilon}{4 e^2}, 
\label{epsilon_s}\\
\frac{1}{\mu_s}&\equiv&\frac{1}{4e^2 \delta \mu}. 
\label{mu_s}
\end{eqnarray}
By using the relations $\frac{e^2}{4\pi\epsilon_0}\simeq 1/137$ and $\epsilon_0 \mu_0=1$ and also the parameters in Table. \ref{parameters}, which are relevant for the honeycomb-layered insulator Na$_2$IrO$_3$\cite{Shitade}, we obtain 
$\epsilon_0/\frac{e^2}{6 \pi |\Delta|d}=0.5$ and $\mu_0 \cdot \frac{e^2v^2}{6 \pi |\Delta|d}=2 \times 10^{-8}$, i.e., $\mu\simeq\mu_0$.
The elastic term for  $a_0^z(=\varphi^z)$ is also induced. 
We can recognize that any potential terms 
(i.e., zeroth-order terms with respect to the derivative $\partial_\mu$) of $A_\mu$ and also $a_\mu^z$ in ${\cal{L}}_{\rm ind}$ 
are absent because of the presence of $U(1)_{\rm em} \times U(1)_z$ gauge symmetry in the fermionic part of the microscopic Lagrangian (\ref{microL}). 
Thus, the low-energy and long-wavelength physics of $A_\mu$ and $a_\mu^z$ 
is described definitely by eq. (\ref{Leff}).

\begin{table}[htdp]
\caption{Parameters used for estimations. These are typical values for 
Na$_2$IrO$_3$,\cite{Shitade} which is a honeycomb-layered insulator with $\lambda_{\rm SO}$ and electron correlation.}
\begin{center}
\begin{tabular}{|c|c|c|c|c|}
\hline
$\Delta$ & $U$ & $d$ & $a$ & $v$
\\ 
\hline
0.5eV & 0.5eV & 10$\AA$ &  10$\AA$&$  3 \times 10^4$m/s 
\\
\hline
\end{tabular}
\label{parameters}
\end{center}
\end{table}%

\section{Magnetic Response}

We consider the static magnetic response. Here, we set $\vec{\bf a}=const.$ 
The equations of motion for spin chemical potential $a_0^z$ (originally, this is the auxiliary field $\varphi^z$) and magnetic field $B$ obtained from eq. (\ref{Leff}) are
\begin{eqnarray}
\epsilon_s \nabla^2 a_0^z+\frac{1}{g}a_0^z&=&\sigma_{xy}^s B, 
\label{poisson+1/g}\\
\frac{1}{\mu} \sum_{j=x,y} \epsilon_{ij} \nabla_j B &=&\sigma_{xy}^s \sum_{j=x,y} \epsilon_{ij} \nabla_j a_0^z.   
\label{Maxwell}
\end{eqnarray}
The {\it r.h.s.} of eqs. (\ref{poisson+1/g}) and (\ref{Maxwell}) are the results from the BF term [see eqs. (\ref{spin-accum}) and (\ref{dualQSHE}) in the Appendix], 
respectively.

We consider a sample in $x \geq 0$, and apply the homogeneous magnetic field $B_0$ parallel to $z$-axis, which is perpendicular to the layers. 
Around the boundary $x=0$, we have the helical edge mode.\cite{KM1,KM2} 
The contribution from the edge mode is discussed in the next section. It will be shown that the edge mode gives a small 
correction via Zeeman coupling. 

Obviously, the fields depend on $x$ only, and the general solution is 
\begin{eqnarray}
\left(
\begin{array}{c}
a_0^z(x)  \\
B(x)  
\end{array}
\right)&=&
\alpha 
\left(
\begin{array}{c}
g\sigma_{xy}^s  \\
1
\end{array}
\right)
+
\nonumber\\
&&
\beta_+ 
\left(
\begin{array}{c}
1/\sigma_{xy}^s\mu \\
1
\end{array}
\right) e^{i k_0 x} 
+ 
\nonumber\\
&&
\beta_- 
\left(
\begin{array}{c}
1/\sigma_{xy}^s\mu \\
1
\end{array}
\right) e^{-i k_0 x},  
\label{general-sol}
\\
k_0&\equiv&{\cal{C}}\sqrt{\frac{1-s}{s}}, 
\label{k0}
\\
{\cal{C}}&\equiv& \sigma_{xy}^s \sqrt{\frac{\mu}{\epsilon_s}}, 
\label{C}
\\
s&\equiv&\sigma_{xy}^{s2} \mu g \geq 0, 
\label{s}
\end{eqnarray} 
where $\alpha$, and $\beta_{\pm}$ are arbitrary constants. The energy functional of the fields is 
\begin{eqnarray}
Ene.=\int d^3x \left\{\frac{1}{2\mu} B^2 + \frac{\sigma_{xy}^2 \mu}{2 {\cal{C}}} ({\bm \nabla} a_0^z)^2 + \frac{1}{2g} a_0^2\right\}, 
\label{ene}
\end{eqnarray}
which will be used to determine the constants. Note that the BF term is absent, since the term does not consume energy.\cite{note}

\subsection{Oscillation effect}
For $s=\sigma_{xy}^{s2}\mu g<1$, $k_0$ is real. 
The Dirichlet-type boundary condition for $B(x)$ at $x=0$ is 
\begin{eqnarray}
\frac{B(0)}{\mu}&=&\frac{B_0}{\mu_0}.  
\label{bc-B}
\end{eqnarray}
We also assume that the fields are real and that the energy functional (\ref{ene}) per period $2\pi/k_0$ along the $x$-direction takes a minimum value. 
Then, we obtain  
\begin{eqnarray}
B(x)&=&B_0 \frac{\mu}{\mu_0} \left\{\frac{1}{1+X} + \frac{X}{1+X} \cos k_0 x\right\}, 
\label{osci-B}
\\
a_0^z(x)&=&B_0 \frac{\mu}{\mu_0}\left\{\frac{g \sigma_{xy}^{s}}{1+X} + \frac{1}{\sigma_{xy}^s \mu}\frac{X}{1+X} \cos k_0 x\right\}, 
\label{osci-a0}
\\
X&=&\frac{2 s^2(1+2 s)}{1+ s^2}. 
\label{X}
\end{eqnarray}
The first term in the {\it r.h.s.} of eq. (\ref{osci-B}) is homogeneous, and the second one shows 
oscillation of the magnetic field. It is reasonable that the amplitude of oscillation vanishes 
in the non-interacting limit $U\rightarrow 0$ [see eqs. (\ref{g}), (\ref{s}), and (\ref{X})]. 

Using the parameters shown in Table \ref{parameters}, we obtain 
$s\simeq 7.0 \times10^{-6}$, and it is hard to observe the oscillation. 
If we have the lattice with $a=350$ nm, 
we obtain $s\simeq 0.99$. In this case, the amplitude of oscillation is about 75 $\%$ of $B_0 \mu / \mu_0$. 
The wavelength of the oscillation is 
\begin{eqnarray}
\lambda_{\rm osci.}=\frac{2\pi}{{\cal{C}}}\sqrt{\frac{s}{1-s}} \simeq 3 \mu {\rm m}. 
\end{eqnarray}
This result is consistent with a long-wavelength approximation since $\lambda_{\rm osci.}\gg a$. 

\subsection{Meissner effect} 
For $s>1$, $k_0$ becomes pure imaginary. We introduce a real value,  
\begin{eqnarray}
\kappa_0 \equiv -i k_0={\cal{C}}\sqrt{\frac{s-1}{s}}.
\label{kappa0}
\end{eqnarray}  
We impose solution (\ref{general-sol}) to be real and finite, and use boundary condition (\ref{bc-B}).
The solution that gives the minimum of energy functional (\ref{ene}) is  
\begin{eqnarray}
B(x)&=&B_0 \frac{\mu}{\mu_0} e^{-\kappa_0 x},  
\label{damp-B}
\\
a_0^z(x)&=&\frac{B_0 }{\sigma_{xy}^s\mu_0} e^{-\kappa_0 x}.
\label{damp-a0}
\end{eqnarray}
It is obvious that the energy of these solutions converges because of the exponential damping. 
The homogeneous part should not appear, since energy functional (\ref{ene}) diverges. 
The above solutions remind us of the Meissner effect with the penetration depth
\begin{eqnarray}
\lambda_{\rm pen.}=\frac{2\pi}{\kappa_0}=\frac{2\pi}{{\cal{C}}}\sqrt{\frac{s}{s-1}}. 
\label{pen}
\end{eqnarray}

It has been pointed out that eqs. (\ref{poisson+1/g}) and (\ref{Maxwell}) lead to  
the London equation for $s\gg1$.\cite{GM}
In this limit, we can neglect the $1/g$ term in the {\it l.h.s.} of eq. (\ref{poisson+1/g}) and 
obtain the London equation by taking the rotation of both sides of eq. (\ref{Maxwell}) and using 
eq. (\ref{poisson+1/g}). 
Amazingly, we obtain the Meissner effect without the London equation in this paper.
Namely, the condition for the Meissner effect is weakened as $s>1$, instead of $s\gg1$.

We can see from eq. (\ref{pen}), large $s$ strengthens the screening. On the other hand, $U$ should 
not be too large, since we are discussing the quantized SH (topological band insulating) phase.\cite{Pesin-Balents,Rachel-LeHur}
Thus, we can see, from eqs.  (\ref{SHC}), (\ref{g}), and (\ref{s}),   
that large $a$ and small $d$ are favorable for observing the Meissner effect. 
We also note that large $\Delta \propto \lambda_{\rm SO}$ shortens the penetration depth, 
since  ${\cal{C}}\propto \Delta^{1/2}$ [see eqs. (\ref{delta_epsilon}),(\ref{C}), and (\ref{pen})].

The physical picture of this Meissner effect is considered to be as follows: the spin-orbit coupling opens 
the topological gap and compensates the energy loss coming from the screening of the magnetic field.

If we have a lattice with $a=380$ nm, instead of $a=10$ $\AA$ in Table \ref{parameters}, we obtain $s \simeq 1.012$ 
and the penetration depth of the magnetic field is estimated to be $\lambda_{\rm pen.}\simeq 2.8$ $\mu$ m. 
This result is consistent with long-wavelength approximation, since $\lambda_{\rm pen.}\gg a$. 

\subsection{Quantized current in topologically non trivial insulators}
The existence of charged current is indicated, since the magnetic field is spatially dependent. 
It may sound curious that current flows in an insulating system. 
We note that the current is a result of a combination of 
spin accumulation (\ref{spin-accum}) and the dual quantized SH effect (\ref{dualQSHE}) derived from the BF term [see Appendix]. 
The spin accumulation causes the spin chemical potential $a_0^z$, and the quantized electric current flows perpendicular to 
$-{\bm \nabla} a_0^z$, i.e., the spin electric field. Therefore, the origin of the current has a direct analog of the quantized Hall effect,\cite{TKNN} in which 
non dissipative quantized transport carried {\it not} by the excited state but by the ground state occurs. 
These are typical transport phenomena in band insulators with non trivial topology\cite{TKNN,Thouless-pumping}.  
We note that such a non dissipative charge transport {\it does not} need the spontaneous $U(1)_{\rm em}$ symmetry breaking. 

In this section, we have considered the bulk state only and the helical edge state was not taken into account. 
We will discuss, in the next section,  
that, as a result of the Zeeman effect, the edge state shows weak diamagnetism independent of the correlation parameter $g$, and gives a 
correction to the results of eqs.  (\ref{osci-B}), (\ref{osci-a0}), (\ref{damp-B}), and (\ref{damp-a0}). 

\section{Contribution from the Helical Edge State} 

In this section, we consider the edge in more detail.  
It is important to estimate the contribution 
from the helical edge mode, which is the hallmark of the quantum SH system\cite{KM1,KM2,HgTe-th,HgTe-exp,Hasan-Kane}. 
 We assume that the edge modes are completely localized at 
the boundary and neglect the broadness. 
First, we neglect Zeeman coupling, and later, we take it into account.

\subsection{Without Zeeman coupling}

In this subsection, we omit Zeeman coupling and show that the helical edge mode has a non-gauge-invariant contribution to the magnetic response. This contribution is cancelled out by the anomalous boundary response as a result of the bulk BF term. 
Namely, only the bulk contribution taken into account in the previous discussion is relevant. 
This anomaly cancellation, renowned as the bulk-edge correspondence, is a natural consequence of the gauge invariance, 
and exactly the reason for the presence of the helical edge mode.\cite{BF1,BF2,BF3,WenEdge,Shizuya,BukkiEdge,Hatsugai1,Hatsugai2}

We consider the layered system in $x \geq 0$, and each layer is normal to the $z$-axis.  
The helical edge mode can be modeled by a pair of quasi one-dimensional (1D) massless Dirac fermions with opposite spin and velocity. 
For simplicity, we assume that edge fermions are localized completely at the boundary. We also assume that the many-body 
interaction between edge fermions is irrelevant, since we are discussing the integral quantized phase.  
We introduce gauge couplings for the fermions in a covariant manner, and integrate out the fermions.
This calculation is a direct analog of that for the chiral edge mode in the quantized Hall effect, and 
can be modeled by a single 1D massless Dirac fermion.\cite{BukkiEdge} 
Then, we obtain the response to the electromagnetic gauge field 
\begin{eqnarray}
&&j^{(edge)z}_\alpha =-\delta(x)\frac{\sigma_{xy}^s}{2} \times
\label{edge-current}\\
&&\left\{g_{\alpha\beta} - (g_{\alpha \alpha^\prime} + 
\epsilon_{\alpha \alpha^\prime})\frac{\partial^{\alpha^\prime}\partial^{\beta^\prime}}{\partial^2}
(g_{\beta^\prime \beta} - \epsilon_{\beta^\prime \beta})\right\}A^\beta, 
\nonumber
\end{eqnarray}
 where $\alpha,\beta=0,y$ denotes the space-time coordinate at the quasi-1D edge lying in the $x=0$ plane, $g^{\alpha\beta}=diag(v^{-2},-1)$, 
 $\partial^2=g_{\alpha\beta} \partial^\alpha \partial^\beta = (\partial_0/v)^2 - \partial_y^2$, and $\epsilon_{\alpha\beta}=-\epsilon_{\beta\alpha}$.  
This current is anomalous since it is not invariant under 
the gauge transformation $A_\alpha \rightarrow A_\alpha + \partial_\alpha \xi$, where $\xi$ is a regular function.  

In the bulk, we have the BF term. The term in the symmetrized form can be written as 
\begin{eqnarray}
{\cal{L}}_{BF}=\frac{\sigma_{xy}^s}{2} \theta(x) \epsilon^{\mu\rho\nu}(a_\mu^z \partial_\rho A_\nu + A_\mu \partial_\rho a_\nu^z).  
\label{BF-boundary}
\end{eqnarray}
We see that the response to the external field $A_\mu$ has a 
boundary term [see the last term]: 
\begin{eqnarray}
j_{\mu}^{(BF)z} &=& \frac{\partial {\cal{L}}_{BF}}{\partial a^{z\mu}}
\label{BF-current}\\
&=&\theta(x) \sigma_{xy}^s \epsilon_{\mu\rho\nu}\partial^\rho A^\nu - \delta(x)\frac{\sigma_{xy}^s}{2}\epsilon_{x\mu\rho}A^\rho.  
\nonumber
\end{eqnarray}
Note that this is spin density ($\mu=0$) or spin current density ($\mu=i$), and 
the boundary term breaks the gauge invariance. 

We consider the response to the static magnetic field. Around the edge $x=0$, we may write 
$A_0=A_x=0$ and $A_y=B(0) x$. Thus, $A_\alpha=\partial_\alpha \left\{B(0) x y\right\}$ 
and from eq. (\ref{edge-current}), we obtain 
\begin{eqnarray}
j_0^{({\rm edge})z}&=&\delta(x)\frac{\sigma_{xy}^s}{2} A_y,
\\
j_y^{({\rm edge})z}&=&0,  
\end{eqnarray}
and from eq. (\ref{BF-current}), 
\begin{eqnarray}
j_0^{({\rm BF})z}&=&\theta(x) \sigma_{xy}^z B - \delta(x)\frac{\sigma_{xy}^s}{2} A_y,  
\\
j_y^{({\rm BF})z}&=&0. 
\end{eqnarray} 
Therefore, the non-gauge-invariant boundary terms cancell each other out. 
Total spin density, which should be inserted in the {\it r.h.s.} of 
eq. (\ref{poisson+1/g}), is  
\begin{eqnarray}
\rho_s=j_0^{({\rm BF})z}+j_0^{({\rm edge})z}=\theta(x)\sigma_{xy}^s B.   
\end{eqnarray}
Thus, we conclude that only 
the bulk contribution is relevant to the magnetic response when we 
neglect Zeeman coupling. 

\subsection{With Zeeman coupling}

We can see that the helical edge mode contributes to the magnetic response as a result of Zeeman coupling.\cite{Kurihara} 
We assume that the Zeeman splitting energy is much less than the bulk band gap $\Delta$. 
As mentioned, the helical edge mode is described by a pair of quasi-1D gapless Dirac fermions.  
When SHC is given by eq. (\ref{SHC}), i.e., quantized as $+1$ with the unit of $e/2\pi$, the edge spectrum is  
\begin{eqnarray}
E_{k_y}^{\uparrow,\downarrow} =\mp v k_y, 
\label{edge_sp}
\end{eqnarray} 
where $k_y$ is the momentum measured from the Fermi points, and 
$-v$ and $+v$ ($v>0$) are velocities for up-spin and down-spin fermions, respectively. In the case of SHC quantized as $-1$, the signs of velocities are opposite to each other.  

As usual metals, Fermi points for up- and down-spin fermions are split by the Zeeman effect, and 
the number density for each spin is generated as 
\begin{eqnarray}
\Delta n^{\uparrow,\downarrow}&=&\pm\left\{\left(\frac{\mu_B B(0)}{v}\right)/\left(\frac{2\pi}{L}\right)\right\}\frac{\delta(x)}{L d} 
\nonumber\\
&=&\pm \frac{\mu_B B(0)}{2 \pi v d}\delta(x),
\label{delta_n}
\end{eqnarray}
where $\mu_B$ is the Bohr magneton, and $L~(\gg a)$ is the total length of the sample edge. Induced spin density is   
\begin{eqnarray}
\Delta \rho_{s}&=&\frac{1}{2}(\Delta n^\uparrow - \Delta n^\downarrow)
\nonumber\\
&=&\frac{\mu_B B(0)}{2 \pi v d}\delta(x). 
\end{eqnarray}

Because of the characteristic feature of the spectrum (\ref{edge_sp}), we have induced current density. For each spin,   
\begin{eqnarray}
\Delta j^{\uparrow,\downarrow}=\pm e v \Delta n^{\uparrow,\downarrow}, 
\end{eqnarray}
and in total, we have   
\begin{eqnarray}
\Delta j&=&\Delta j^{\uparrow}+\Delta j^{\downarrow}
\nonumber\\
&=& \frac{e \mu_B}{\pi d} B(0)\delta(x).  
\label{edge_current}
\end{eqnarray}
Therefore, the equations of motion (\ref{poisson+1/g}) and (\ref{Maxwell}) are modified as 
\begin{eqnarray}
\epsilon_s \frac{d^2 a_0^z}{d x^2} +\frac{a_0^z}{g}&=&\sigma_{xy}^s B+\frac{\mu_B B(0)}{2 \pi v}\delta(x), 
\label{poisson+1/g+edge}\\
\frac{1}{\mu} \frac{d B}{d x}&=&\sigma_{xy}^s \frac{d a_0^z}{d x} -\frac{e \mu_B B(0)}{\pi d} \delta(x).   
\label{Maxwell+edge}
\end{eqnarray}
Instead of eq. (\ref{bc-B}), the Dirichlet boundary condition for $B$ at $x=0$ is  
\begin{eqnarray}
\frac{B(0)}{\mu}&=&\frac{B_0}{\mu_0}-\frac{e\mu_B}{\pi d} B(0), 
\label{bc-B+edge}
\end{eqnarray}
i.e., the value $B(0)$ is shifted. Thus, the corrected results are given by replacing the coefficient $B_0$ in eqs. (\ref{osci-B}),  (\ref{osci-a0}), (\ref{damp-B}), and (\ref{damp-a0}) as 
\begin{eqnarray}
B_0 \rightarrow \frac{B_0}{1+ (e \mu \mu_B/\pi d)}.  
\end{eqnarray}
We note that $e \mu \mu_B / \pi d \simeq 5.6 \times 10^{-6}$ for $\mu=\mu_0$ and $d=10\AA$.  

The result indicates that the helical edge mode shows the weak diamagnetism independent of the parameter $g$. 

\section{The Electric Conductivity: Comparision with Superconductivity} 

Finally, we examine the electric conductivity in our system and compare it with the superconductivity. 
We integrate out $\vec{a}_0$ from the effective Lagrangian (\ref{Leff}). 
Note that this integration is  justified since $\vec{a}_0$ was originally introduced as the auxiliary field of the Stratonovich-Hubbard transformation. 
We do not integrate out $\vec{\bf a}$, since this field was introduced merely to estimate the spin current. 
This integration is straightforward, since the Lagrangian (\ref{Leff}) is  quadratic with respect to $\vec{a}_0$. We obtain 
\begin{eqnarray}
{\cal{L}}_{\rm eff}^\prime&=&\sigma_{xy}^{s}\sum_{ij} \epsilon_{ij} a_i^z E_j-
\frac{\sigma_{xy}^{s 2}}{2} {\bm A}^T\frac{{\nabla^2}}{\epsilon_s{\nabla^2}+g^{-1}}{\bm A}^T 
\nonumber\\
&&+\frac{\epsilon E^2 }{2}- \frac{B^2}{2\mu} 
\nonumber\\
&&+ ({\rm terms~independent~of}~a_i^z~{\rm and}~A_\mu), 
\end{eqnarray}
where $A_i^T=\sum_j [\delta_{ij} -\partial_i \partial_j/\nabla^2]A_j$ is the 
transverse (i.e., gauge invariant) component of $A_i$. 
We set $\vec{\bm a}$ to be constant after the integration. The quantized SH current is obtained from the first term, as expected. The electric current is\cite{note_J} 
\begin{eqnarray}
{\bm J}=-\frac{\sigma_{xy}^{s 2}{\nabla^2}}{\epsilon_s{\nabla^2}+ g^{-1}}{\bm A}^T.  
\label{eff-current}
\end{eqnarray}
We may take the gauge $A_0={\bm \nabla} \cdot {\bm A}=0$ then we have ${\bm E}=-\dot{\bm A}$ and ${\bm A}^T={\bm A}$. 
Thus the electric conductivity is  
\begin{eqnarray}
\sigma_{xx}(\omega, {\bm {q}})=-\frac{\sigma_{xy}^{s 2}{\bm q}^2}
{\epsilon_s{\bm q}^2 - g^{-1}} \frac{1}{i \omega}, 
\label{conductivity}
\end{eqnarray}
which vanishes in the DC limit taking ${\bm q}\rightarrow 0$ first and $\omega \rightarrow 0$ later when $1/g \neq 0$. \cite{Mahan}
Namely, the system is insulating. 
Then, we conclude that the system shows the Meissner effect without infinite DC conductivity when $g>(\sigma_{xy}^{s2} \mu)^{-1}$,  
in contrast to the superconductivity.  A similar result has been obtained by the Maxwell-Chern-Simons (MCS) theory, which is the low-energy and long-wavelength effective theory for the time-reversal-violating topological band insulator, i.e., the quantized Hall system.\cite{M-C-S1,M-C-S2,M-C-S3,Kotetes-Varelogiannis}

Eqs. (\ref{eff-current}) and (\ref{conductivity}) indicate that 
the system becomes superconducting at the point $1/g=0$ [see also refs. 14, 33]. Actually, we can see the infinite DC conductivity from the Kramers-Kr{\"o}nig relation as 
${\rm Re}[ \sigma_{xx} (\omega,0)]\propto \delta (\omega)$. Unfortunately, this point 
is difficult to realize in the quantized SH phase, as mentioned in Introduction.  

\section{Summary}

In this study, we investigated the magnetic response in the quantized SH phase of a layered-honeycomb lattice system 
with the intrinsic spin-orbit coupling $\lambda_{\rm SO}$ and on-site Hubbard $U$. 
When $g\equiv 4 U a^2 d / 3< (\sigma_{xy}^{s2}  \mu)^{-1}$, where $a$ and $d$ are the lattice constant and interlayer distance, respectively, 
the magnetic field inside the sample oscillates spatially around a constant value. 
The oscillation vanishes in the non-interacting limit $U\rightarrow 0$. 
When $g>(\sigma_{xy}^{s2} \mu)^{-1}$, the Meissner effect occurs. 
It may be possible to see the oscillation or Meissner effect in a superlattice system with an appropriately large $a$.
As a result of Zeeman coupling, the helical edge state\cite{KM1,KM2} shows the weak diamagnetism that is independent of $g$.

\section*{Acknowledgment}
The authors are grateful to T. Fujii, N. Furukawa, N. Hatano, D. S. Hirashima, K.-I. Imura, T. Kato, S. Kurihara, S. Miyashita, 
T. Oka, Masahiro Sato, Masatoshi Sato, A. Shitade, and K. Ueda for fruitful discussions and stimulating comments. 
J.G. is financially supported by a Grant-in-Aid for Scientific Research 
from Japan Society for the Promotion of Science, Grant 
No. 18540381, as well as by Core Research for Evolutional
Science and Technology (CREST) of Japan Science and Technology Agency. 

\appendix

\section{Physical implications of BF term}

In this Appendix, we summarized the properties of the BF term\cite{GM,GMI,BF1,BF2,BF3,BF4} in eq. (\ref{Leff}) 
 \begin{eqnarray}
{\cal{L}}_{\rm BF}= \sigma_{xy}^{s} \epsilon^{\mu\rho\mu}a_\mu \partial_\rho A_\nu,  
\label{BF}
 \end{eqnarray}
where $\sigma_{xy}^s$ is the quantized SHC given by eq. (\ref{SHC}). 
 
\subsection{Quantized SH effect}
The spin current density obtained from the term is 
\begin{eqnarray}
j_i^{s}=\frac{\partial {\cal{L}}_{\rm BF}}{\partial a_i^z}=\sigma_{xy}^s \sum_{j=x,y} \epsilon_{ij} E_j.  
\label{QSHE}
\end{eqnarray}
This shows the quantized SH effect. 

\subsection{Spin accumulation}
The spin density is 
\begin{eqnarray}
\rho^{s}=\frac{\partial {\cal{L}}_{\rm BF}}{\partial a_0^z}=\sigma_{xy}^s B, 
\label{spin-accum}
\end{eqnarray}
where $B$ is the magnetic field perpendicular to the honeycomb lattice layers. 
It resembles the Zeeman effect, but an essential difference is that the coefficient is not the Bohr magneton but the quantized SHC. 

\subsection{Dual quantized SH effect}
The electric current density is  
\begin{eqnarray}
j_i= \frac{\partial {\cal{L}}_{\rm BF}}{\partial A_i}=\sigma_{xy}^s\sum_{j=x,y}\epsilon_{ij} \nabla_j a_0^z.    
\label{dualQSHE}
\end{eqnarray}
This shows that the current flows perpendicular 
to the gradient of spin chemical potential $a_0^z$, namely, the spin electric field. 
This may be called the dual quantized SH effect.


\begin{thebibliography}{99}
\bibitem{note2} According to ref. 7, we distinguish ``quantized" and ``quantum" spin Hall effects as follows. 
In the quantized spin Hall effect, electron spin $S_z$ is conserved. Thus, the spin Hall conductivity is well defined and 
shows quantization. In quantum spin Hall effect $S_z$-conservation is violated by time-reversal-invariant perturbations and the conductivity 
is no longer well defined, but the helical edge state, which is a hallmark of the quantum SH effect,  still survives robustly.\cite{KM1,KM2,Hasan-Kane}  
As is mentioned in the text, we focus on the quantized spin Hall phase.
\bibitem{note0}
It has been pointed out by Rachel and Le Hur \cite{Rachel-LeHur} that, in the layered-honeycomb lattice system, 
the phase transition occurs around $U/t\simeq 3$ when $\lambda_{\rm SO}/t\simeq1$ ($t$; the hopping energy).
\bibitem{note} 
Hereafter, we use the Lagrange formalism. The reason is that the topological BF term, which plays 
a crucial role in our discussion, does not appear in the Hamilton formalism explicitly and the formulation becomes complicated\cite{BF1,BF2,BF3}.
We can explain this as follows: The quantized SH effect is one of the physical implications of the term [see Appendix].  
It is renown that there is no energy consumption in the quantized SH effect, like the quantized charge Hall effect, 
since current response is perpendicular to the external field.
\bibitem{note_J} Substituting eq. (\ref{eff-current}) into the Maxwell equation ${\bm \nabla} \times {\bm B}/\mu=-{\bm J}$, where ${\bm B}=\hat{\bm z} B$, and taking the rotation of both sides of 
the equation, we have 
$$
\frac{1}{\mu}\nabla \times \nabla \times {\bm B}=-\sigma_{xy}^{s2}\frac{\nabla^2}{\epsilon_s \nabla^2 +  g^{-1}} {\bm B}. 
$$
To solve it, we can obtain the equivalent results discussed in $\S$ 4.
\end{thebibliography}
\end{document}